\documentstyle[floats,aps]{revtex}
\begin{document}
\input epsf

  \font\twelvemib=cmmib10 scaled 1200
  \font\elevenmib=cmmib10 scaled 1095
  \font\tenmib=cmmib10
  \font\eightmib=cmmib10 scaled 800
  \font\sixmib=cmmib10 scaled 667
  \skewchar\elevenmib='177
  \newfam\mibfam
  \def\mib{\fam\mibfam\tenmib}
  \textfont\mibfam=\tenmib
  \scriptfont\mibfam=\eightmib
  \scriptscriptfont\mibfam=\sixmib

\draft

\twocolumn[\hsize\textwidth\columnwidth\hsize\csname  
@twocolumnfalse\endcsname

\title{The Interacting Impurity Josephson Junction: Variational Wavefunctions
and Slave Boson Mean Field Theory}
\author{A. V. Rozhkov and Daniel P. Arovas}
\address{Department of Physics, University of California at San Diego,
La Jolla CA 92093}
\date{\today}

\maketitle

\begin{abstract}
We investigate the Josephson coupling between two superconductors
mediated through an infinite $U$ Anderson impurity, adapting a variational
wavefunction approach which has proved successful for the Kondo model.
Unlike the Kondo problem, however, a crossing of singlet and
doublet state energies may be produced by varying the ratio of Kondo
energy to superconducting gap, in agreement with recent work of
Clerk and Ambegaokar.  We construct the phase diagram for the junction
and discuss properties of different phases.  In addition, we find the
singlet and doublet state energies within a slave boson mean field approach.
We find the slave boson mean field treatment is unable to account for the
level crossing.
\end{abstract}

\vskip2pc]

\narrowtext

\section{Introduction}
The Ambegaokar-Baratoff formula, $I_{\rm c}=\pi\Delta/2eR$, relates the
critical current in a Josephson junction to the superconducting gap $\Delta$
and the normal state junction resistance $R$.  This result is perturbative
in the electron tunneling amplitude $t$; the normal state conductance is
proportional to $|t|^2$ when $t$ is small.  In certain instances, however,
it may be that the tunneling is mediated through a magnetic impurity,
rather than taking place directly from one superconductor to the other.
This situation has been considered by a number of authors
\cite{kulik,SS,BKS,GM,SK,RA,CA}.  The principal result is that magnetic
impurity-mediated tunneling results in a {\it negative\/} contribution to
$I_{\rm c}$.  As Kulik originally argued \cite{kulik}, the magnetic impurity
gives rise to an effective spin-flip hopping amplitude $t_{\rm sf}$ between
the superconductors.  Since spin-flip tunneling results in a sign change of the
Cooper pair singlet ($ \uparrow \downarrow$ to $ \downarrow \uparrow$)
\cite{SK}, the critical current
is $I_{\rm c}\propto |t|^2-|t_{\rm sf}|^2$.  When $I_{\rm c}<0$, one has
a $\pi$-junction, for which the ground state energy is minimized when the phase
difference between the superconductors is $\delta=\pi$.  Such $\pi$-junctions
break time-reversal symmetry ($ {\cal T}$), and a ring containing a single
$\pi$-junction will enclose trapped flux \cite{BKS}.

This analysis suffices in the perturbative limit where $t$ is small.
If the impurity level energy is $  \varepsilon ^{\vphantom{\dagger}}_0$ and
the Coulomb integral is $U$,
the condition for a magnetic ground state (and a $\pi$-junction)
is $U>-  \varepsilon ^{\vphantom{\dagger}}_0>0$ \cite{SK}.  In the
nonperturbative regime, a new energy
scale arises: the bare impurity level width $\Gamma\sim \pi\rho |t|^2$,
where $t$ is the electrode-impurity hopping matrix element and $\rho$ the
electrode density of states.  The $T=0$ phase diagram as a function of
$-\varepsilon ^{\vphantom{\dagger}}_0/\Delta$, $U/\Delta$, and $\Gamma/\Delta$
was investigated by the authors
\cite{RA} within the Hartree-Fock (HF) approximation, where it was found
that $\pi$-junction behavior occurs for
$U>-\varepsilon ^{\vphantom{\dagger}}_0>0$, provided $\Gamma$ is
sufficiently small
($\Gamma \,{\raise.3ex\hbox{$<$\kern-.75em\lower1ex\hbox{$\sim$}}}\, U$).
However, when $\Delta=0$, it is known
that the HF approximation is unable to describe the formation of
a Kondo singlet at energy scales below $T ^{\vphantom{\dagger}}_ {\rm K}\equiv
W\exp(-\pi |  \varepsilon ^{\vphantom{\dagger}}_0|/2\Gamma)$, where $W$ is the
half-bandwidth in the electrodes.
In our problem, then, one expects that the Kondo effect will be mitigated
whenever $\Delta \,{\raise.3ex\hbox{$>$\kern-.75em\lower1ex\hbox{$\sim$}}}\,
T^{\vphantom{\dagger}}_{\rm K}$.
Roughly speaking, if $\Delta>T_{\rm K}$, the ground state of the system
is a Kramers doublet, and breaks time reversal symmetry, whereas if
$T ^{\vphantom{\dagger}}_ {\rm K}>\Delta$, the ground state is a hybrid
singlet formed from
electrons on the impurity and in the superconducting electrodes \cite{GM}.

Recently, Clerk and Ambegaokar \cite{CA} applied a generalization of
the non-crossing approximation (NCA), a partial summation scheme, to
attack this problem in the $U\to\infty$ limit \cite{CA}.
Adaptation of this method to the interacting impurity Josephson junction
allows one to see a transition from $0$-junction to $\pi$-junction when the
superconducting gap becomes comparable to the Kondo temperature.

In this paper, we also explore the $U\to\infty$ limit, using two approaches.
The first is a variational wave function calculation, similar to that
used by Varma and Yafet in the Kondo problem \cite{VY,GS}.  We generalize
this wave function in two respects.  Firstly, there are two
superconductors connected to the impurity.  This is in contrast with the usual
setup of Kondo problem where we have only metallic electrode. Secondly, the
spin-$ \frac{1}{2}$ state must be considered as well.  We find, in agreement
with ref. \cite{CA}, that a first order transition occurs at
$T ^{\vphantom{\dagger}}_ {\rm K}/\Delta
\approx 1$.  We also show how this transition may be precipitated by a change
in the phase difference $\delta$, as we found in ref. \cite{RA}.

The second calculation we describe is the slave boson mean field theory.  While
the this method does confirm that the singlet state gives a $0$-junction and the
Kramers doublet a $\pi$-junction, within the static slave boson mean field
solution the singlet is always lower in energy (or the states are degenerate).
The NCA, which goes beyond the mean field level, is able to describe the
transition.

\section{Variational Wavefunction Approach}
We start with the grand canonical Hamiltonian $ {\cal K}\equiv {\cal H}-\mu N$,
\begin{eqnarray}
{\cal K}&=&\sum_\alpha\sum_ {\mib q}\biggl\{
 \xi ^{\vphantom{\dagger}}_{ {\mib q}\alpha}\Big(
\psi ^\dagger_{ {\mib q}\alpha \uparrow }
\psi ^{\vphantom{\dagger}}_{ {\mib q}\alpha \uparrow }+
\psi ^\dagger_{ {\mib q}\alpha \downarrow}
\psi ^{\vphantom{\dagger}}_{ {\mib q}\alpha \downarrow}\Big)
\nonumber\\
&&+\Delta ^{\vphantom{\dagger}}_\alpha\Big( e^{i\delta_\alpha}\,
\psi ^\dagger_{ {\mib q}\alpha \uparrow }
\psi ^\dagger_{- {\mib q}\alpha \downarrow} + e^{-i\delta_\alpha}\,
\psi ^{\vphantom{\dagger}}_{- {\mib q}\alpha \downarrow}
\psi ^{\vphantom{\dagger}}_{ {\mib q}\alpha \uparrow }\Big)
\nonumber\\
&&-{1\over \sqrt{ {\cal N}_\alpha}}\sum_{\sigma= \uparrow, \downarrow}
\Big(t ^{\vphantom{\dagger}}_\alpha\,
\psi ^\dagger_{ {\mib q}\alpha\sigma}c ^{\vphantom{\dagger}}_\sigma +
t^*_\alpha\, c ^\dagger_\sigma
\psi ^{\vphantom{\dagger}}_{ {\mib q}\alpha\sigma}\Big)\biggr\}\nonumber\\
&&+  \varepsilon ^{\vphantom{\dagger}}_0\Big( c ^\dagger_ \uparrow
c ^{\vphantom{\dagger}}_ \uparrow + c ^\dagger_ \downarrow 
c ^{\vphantom{\dagger}}_ \downarrow \Big)
+U c ^\dagger_ \uparrow  c ^\dagger_ \downarrow
c ^{\vphantom{\dagger}}_ \downarrow  c ^{\vphantom{\dagger}}_ \uparrow \ ,
\label{hamil}
\end{eqnarray}
where $\alpha$ labels the electrode, $\xi_{ {\mib q}\alpha}$ is the dispersion
in the $\alpha$ electrode relative to the chemical potential, $\Delta_\alpha$
is the modulus of the superconducting gap and $\delta_\alpha$ its phase,
$ {\cal N}_\alpha$ is the number of unit cells in electrode $\alpha$,
$t_\alpha$ is the hopping amplitude from electrode $\alpha$ to the impurity,
and $  \varepsilon ^{\vphantom{\dagger}}_0$ is the bare impurity energy.
We set $U\to\infty$, which leads to the constraint
$\sum_\sigma c ^\dagger_\sigma c ^{\vphantom{\dagger}}_\sigma \le 1$.

Defining the angle $\theta_{ {\mib q}\alpha}\equiv\tan^{-1}(\Delta_\alpha/
\xi ^{\vphantom{\dagger}}_{ {\mib q}\alpha})$ and the usual BCS coherence
factors $u^{\vphantom{\dagger}}_{ {\mib q} \alpha}=\cos( \frac{1}{2}
\theta^{\vphantom{\dagger}}_{ {\mib q}\alpha})$,
$ v ^{\vphantom{\dagger}}_{ {\mib q} \alpha}=\sin( \frac{1}{2}
\theta ^{\vphantom{\dagger}}_{ {\mib q}\alpha})\exp(i\delta_\alpha)$,
we express $ {\cal K}= {\cal K}_0+ {\cal K}_1$
in terms of the Bogoliubov quasiparticle operators:
\begin{eqnarray*}
{\cal K}_0&=&\sum_{ {\mib q},\alpha} E ^{\vphantom{\dagger}}_{{\mib q}\alpha}\,
(\gamma^\dagger_{{\mib q}\alpha\uparrow}\gamma^{\vphantom{\dagger}}_{{\mib q}
\alpha\uparrow} + \gamma^\dagger_{{\mib q}\alpha\downarrow}
\gamma^{\vphantom{\dagger}}_{ {\mib q} \alpha  \downarrow })
+\varepsilon^{\vphantom{\dagger}}_0 ( c ^\dagger_ \uparrow
c ^{\vphantom{\dagger}}_ \uparrow  + c ^\dagger_ \downarrow
c ^{\vphantom{\dagger}}_ \downarrow )\\
&&\quad+U c ^\dagger_ \uparrow   c ^\dagger_ \downarrow
c ^{\vphantom{\dagger}}_ \downarrow   c ^{\vphantom{\dagger}}_ \uparrow \\
{\cal K}_1&=&\sum_{ {\mib q},\alpha} t ^{\vphantom{\dagger}}_{ {\mib q} \alpha}
\biggl\{ u ^{\vphantom{\dagger}}_{ {\mib q} \alpha} ( \gamma ^\dagger_{
{\mib q} \alpha  \uparrow }  c ^{\vphantom{\dagger}}_ \uparrow +
\gamma ^\dagger_{ {\mib q} \alpha  \downarrow }
c ^{\vphantom{\dagger}}_ \downarrow + c ^\dagger_ \uparrow 
\gamma ^{\vphantom{\dagger}}_{ {\mib q} \alpha  \uparrow }+
c ^\dagger_ \downarrow   \gamma ^{\vphantom{\dagger}}_{ {\mib q} \alpha 
\downarrow })\\
&&+ v ^{\vphantom{\dagger}}_{ {\mib q} \alpha} ( \gamma ^\dagger_{ {\mib q}
\alpha  \uparrow }  c ^\dagger_ \downarrow - \gamma ^\dagger_{ {\mib q}
\alpha  \downarrow }  c ^\dagger_ \uparrow )+ v^*_{ {\mib q} \alpha}
( \gamma ^{\vphantom{\dagger}}_{ {\mib q} \alpha  \downarrow } 
c ^{\vphantom{\dagger}}_ \uparrow - \gamma ^{\vphantom{\dagger}}_{ {\mib q}
\alpha  \uparrow }  c ^{\vphantom{\dagger}}_ \downarrow )\biggr\}\ ,
\end{eqnarray*}
where $E ^{\vphantom{\dagger}}_{ {\mib q}\alpha}=\sqrt{\xi^2_{
{\mib q}\alpha}+\Delta_\alpha^2}$, and
$ t ^{\vphantom{\dagger}}_{ {\mib q} \alpha}\equiv |t ^{\vphantom{\dagger}}_\alpha|/\sqrt{ {\cal N}_\alpha}$.

We now  two variational many-body states for the $U=\infty$ limit:
a singlet,
\begin{eqnarray}
|\,{\rm S}\,\rangle&\equiv&\biggl\{ A+\sum_{ {\mib q},\alpha}\frac{1}{\sqrt{2}}
B ^{\vphantom{\dagger}}_{ {\mib q}\alpha}( \gamma ^\dagger_{ {\mib q} \alpha
\uparrow } c ^\dagger_ \downarrow - \gamma ^\dagger_{ {\mib q} \alpha
\downarrow } c ^\dagger_ \uparrow )\nonumber\\
&&\qquad+\sum_{ {\mib q},\alpha\atop {\mib q}',\alpha'}
C^{\alpha\alpha'}_{ {\mib q} {\mib q}'}  \gamma ^\dagger_{ {\mib q} \alpha
\uparrow}\gamma^\dagger_{{\mib q}' \alpha'\downarrow }\biggr\}|\,0\,\rangle\ ,
\end{eqnarray}
and a doublet,
\begin{eqnarray}
&&|\,{\rm D} \uparrow\,\rangle\equiv\biggl\{ {\tilde A} c ^\dagger_ \uparrow
+\sum_{ {\mib q},\alpha}{\tilde B} ^{\vphantom{\dagger}}_{ {\mib q}\alpha}
\gamma ^\dagger_{ {\mib q} \alpha  \uparrow }+\sum_{ {\mib q},\alpha\atop
{\mib q}',\alpha'}\Bigl[ {\tilde C}^{\alpha\alpha'}_{ {\mib q} {\mib q}'}
\gamma ^\dagger_{ {\mib q} \alpha  \uparrow } \gamma ^\dagger_{ {\mib q}'
\alpha'  \downarrow } c ^\dagger_ \uparrow \nonumber\\
&&\ +\frac{1}{\sqrt{3}}{\tilde D}^{\alpha\alpha'}_{ {\mib q} {\mib q}'}
( \gamma ^\dagger_{ {\mib q} \alpha  \uparrow } \gamma ^\dagger_{ {\mib q}'
\alpha'  \downarrow } c ^\dagger_ \uparrow - \gamma ^\dagger_{ {\mib q} \alpha
\uparrow } \gamma ^\dagger_{ {\mib q}' \alpha'  \uparrow }
c ^\dagger_ \downarrow )\Bigr]\biggr\}|\,0\,\rangle\ ,
\end{eqnarray}
where $|\,0\,\rangle$ is the fermion vacuum (the other doublet state
$|\,{\rm D} \downarrow\,\rangle$
is obtained by rotating the spins by $\pi$ about the $y$-axis).  Here,
$C^{\alpha'\alpha}_{{\mib q}'{\mib q}}=C^{\alpha\alpha'}_{{\mib q}{\mib q}'}$,
${\tilde C}^{\alpha'\alpha}_{ {\mib q}' {\mib q}}
={\tilde C}^{\alpha\alpha'}_{ {\mib q} {\mib q}'}$, and
${\tilde D}^{\alpha'\alpha}_{ {\mib q}' {\mib q}}=
-{\tilde D}^{\alpha\alpha'}_{ {\mib q} {\mib q}'}$.
We next set to zero the variations
\begin{equation}
\delta\langle\,\Psi\,| {\cal K}_0+ {\cal K}_1|\,\Psi\,\rangle-E\,\delta\langle
\,\Psi\,|\,\Psi\,\rangle=0\ ,
\end{equation}
where $|\,\Psi\,\rangle$ is $|\,{\rm S}\,\rangle$ or
$|\,{\rm D} \uparrow\,\rangle$, to obtain equations
relating the variational coefficients.  We find that $A$ and the matrix
$C^{\alpha\alpha'}_{ {\mib q} {\mib q}'}$ may be expressed in terms of the
coefficients $B ^{\vphantom{\dagger}}_{ {\mib q}\alpha}$,
and similarly ${\tilde A}$ and the
matrices ${\tilde C}^{\alpha\alpha'}_{ {\mib q} {\mib q}'}$ and
${\tilde D}^{\alpha\alpha'}_{ {\mib q} {\mib q}'}$ may be expressed in terms
of the coefficients ${\tilde B} ^{\vphantom{\dagger}}_{ {\mib q}\alpha}$.
We then obtain the two eigenvalue equations for the singlet and doublet
energies $E$ and ${\tilde E}$, respectively: 
\begin{eqnarray}
\sum_{ {\mib q}'\alpha'}&&\biggl[{2\, v ^{\vphantom{\dagger}}_{ {\mib q}\alpha}
\, v^*_{ {\mib q}'\alpha'}\, t ^{\vphantom{\dagger}}_{ {\mib q} \alpha}\,
t ^{\vphantom{\dagger}}_{ {\mib q}'\alpha'}\over E}+
{ u^{\vphantom{\dagger}}_{ {\mib q} \alpha}\,
u ^{\vphantom{\dagger}}_{ {\mib q}'\alpha'}\,
t ^{\vphantom{\dagger}}_{ {\mib q} \alpha}\,
t_{ {\mib q}'\alpha'}\over E-E ^{\vphantom{\dagger}}_{ {\mib q}\alpha}-
E ^{\vphantom{\dagger}}_{ {\mib q}'\alpha'}}\biggr]
B ^{\vphantom{\dagger}}_{ {\mib q}'\alpha'}\cr
&&=\biggl[E-E ^{\vphantom{\dagger}}_{ {\mib q}\alpha}-
\varepsilon ^{\vphantom{\dagger}}_0-\sum_{ {\mib q}',\alpha'}
{u^2_{ {\mib q}'\alpha'}\, t^2_{ {\mib q}'\alpha'}\over
E-E ^{\vphantom{\dagger}}_{ {\mib q}\alpha}-
E ^{\vphantom{\dagger}}_{ {\mib q}'\alpha'}}\biggr]
B ^{\vphantom{\dagger}}_{ {\mib q}\alpha}
\end{eqnarray}
and
\begin{eqnarray}
\sum_{ {\mib q}'\alpha'}&&\biggl[{ u ^{\vphantom{\dagger}}_{ {\mib q} \alpha}
\, u ^{\vphantom{\dagger}}_{ {\mib q}'\alpha'}\,
t ^{\vphantom{\dagger}}_{ {\mib q} \alpha}\,
t ^{\vphantom{\dagger}}_{ {\mib q}'\alpha'}\over {\tilde E}-
 \varepsilon ^{\vphantom{\dagger}}_0}+{ v ^{\vphantom{\dagger}}_{ {\mib q}
\alpha}\,v^*_{ {\mib q}'\alpha'}\, t ^{\vphantom{\dagger}}_{ {\mib q} \alpha}\,
t_{ {\mib q}'\alpha'}\over{\tilde E}-  \varepsilon ^{\vphantom{\dagger}}_0-
E ^{\vphantom{\dagger}}_{ {\mib q}\alpha}-
E ^{\vphantom{\dagger}}_{ {\mib q}'\alpha'}}
\biggr]{\tilde B} ^{\vphantom{\dagger}}_{ {\mib q}'\alpha'}\cr
&&=\biggl[{\tilde E}-E ^{\vphantom{\dagger}}_{ {\mib q}\alpha}-
\sum_{ {\mib q}',\alpha'}
{2|v_{ {\mib q}'\alpha'}|^2\,t^2_{ {\mib q}'\alpha'}\over{\tilde E}-
 \varepsilon ^{\vphantom{\dagger}}_0-
E ^{\vphantom{\dagger}}_{ {\mib q}\alpha}-
E ^{\vphantom{\dagger}}_{ {\mib q}'\alpha'}}\biggr]
{\tilde B} ^{\vphantom{\dagger}}_{ {\mib q}\alpha}
\end{eqnarray}

\begin{figure} [!t]
\centering
\leavevmode
\epsfxsize=8cm
\epsfysize=8cm
\epsfbox[18 144 592 718] {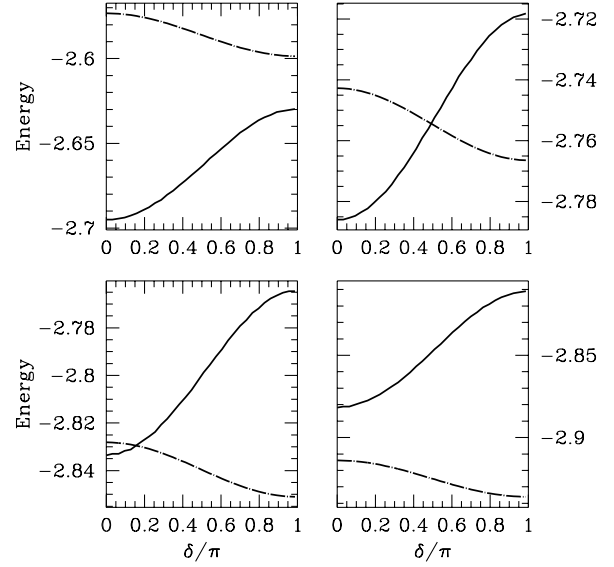}
\caption[]
{\label{afig} Singlet (solid) and doublet (dot-dash) energies {\it versus\/}
phase difference for $  \varepsilon ^{\vphantom{\dagger}}_0=-1.6$ (upper left),
$\varepsilon ^{\vphantom{\dagger}}_0=-1.8$ (upper right),
$\varepsilon ^{\vphantom{\dagger}}_0=-1.9$ (lower left), and
$\varepsilon ^{\vphantom{\dagger}}_0=-2.0$ (lower right).  In all cases
$\Delta=1$ and $ \Gamma ^{\vphantom{\dagger}}_ {\rm L}=
\Gamma ^{\vphantom{\dagger}}_ {\rm R}=0.3\pi$.}
\end{figure}

\begin{figure} [!t]
\centering
\leavevmode
\epsfxsize=8cm
\epsfysize=8cm
\epsfbox[18 144 592 718] {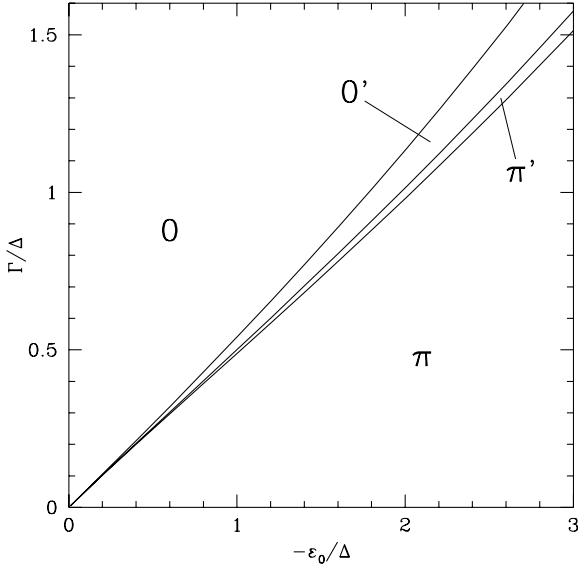}
\caption[]
{\label{bfig} Variational wavefunction phase diagram for the symmetric
case $ \Gamma ^{\vphantom{\dagger}}_ {\rm L}=
\Gamma ^{\vphantom{\dagger}}_ {\rm R}\equiv\Gamma$.}
\end{figure}

We solve these equations numerically for the symmetric case
$\Delta ^{\vphantom{\dagger}}_ {\rm L}=\Delta ^{\vphantom{\dagger}}_ {\rm R}
=\Delta$, $t ^{\vphantom{\dagger}}_ {\rm L}=t^{\vphantom{\dagger}}_{\rm R}=t$,
$\Gamma ^{\vphantom{\dagger}}_ {\rm L}=\Gamma ^{\vphantom{\dagger}}_ {\rm R}
=\Gamma$.  The normal state of each
electrode is described by a flat band of width $2W$; we use $W/\Delta=10$
in our calculations, but the general features are rather insensitive to
the value of $W$.  Energy {\it versus\/} phase difference, $E(\delta$), is
plotted in figure \ref{afig} for $\Delta=1$, $\Gamma=0.3\pi$, for four different
values of $-  \varepsilon ^{\vphantom{\dagger}}_0$.  When
$\varepsilon ^{\vphantom{\dagger}}_0=-1.6$, the singlet state is the ground
state for all $\delta$, while for $  \varepsilon ^{\vphantom{\dagger}}_0=-2.0$,
the ground state is always the doublet.
These are $0$- and $\pi$-junctions, respectively.  For
$\varepsilon ^{\vphantom{\dagger}}_0=-1.8$, the
curves cross, and the ground state energy has a kink as a function of
$\delta$.  Both $\delta=0$ and $\delta=\pi$ are local minima in $E(\delta)$,
with $\delta=0$ the global minimum.  Using the terminology of ref. \cite{RA},
this is a $0'$-junction.  The final case of the $\pi'$-junction is reflected
in the curves for $  \varepsilon ^{\vphantom{\dagger}}_0=-1.9$, where
$\delta=0$ is a local minimum and
$\delta=\pi$ the global minimum of the energy.

The phase diagram is displayed in fig. 2.  In the HF treatment
\cite{RA}, one also finds a transition from $\pi$ to $\pi'$ to $0'$ to
$0$-junction phases as $\Gamma$ is increased from zero.  However, the critical
value of $\Gamma$ there turns out to be inversely proportional to
$\ln U$.  There are two reasons for this: first, the HF calculations
assumed an infinite bandwidth, and second, HF is unable to describe the
formation of a Kondo singlet.  Here, the
transition from $0$-junction to $\pi$-junction occurs roughly
for $\Delta \sim T ^{\vphantom{\dagger}}_ {\rm K}$, in agreement with
ref. \cite{CA}.  Indeed,
when $0<-  \varepsilon ^{\vphantom{\dagger}}_0,\Gamma\ll\Delta$ this relation
can be derived within
perturbation theory. The perturbative value of energy of the singlet state is
\begin{equation}
E_{0}\approx -{4\Gamma\over\pi}\ln{2W\over\Delta},
\end{equation}
while the energy of the doublet is
\begin{equation}
E_{\uparrow }\approx   \varepsilon ^{\vphantom{\dagger}}_0-{2\Gamma\over\pi}
\ln {2W\over\Delta}.
\end{equation}
Comparing these two we find that the symmetry of the ground state
changes when
\begin{equation}
\Delta =2W\exp (-\pi |  \varepsilon ^{\vphantom{\dagger}}_0|/2\Gamma )=
2T ^{\vphantom{\dagger}}_ {\rm K}\ .
\end{equation}
This predicts a straight line for $\Gamma$ {\it versus\/}
$-  \varepsilon ^{\vphantom{\dagger}}_0$ with
a slope $\Gamma/(-  \varepsilon ^{\vphantom{\dagger}}_0)=\pi/2\ln(2W/\Delta)$.
With $W=10\Delta$, the
slope is $0.524$, and the line would be bounded by the $0-0'$ and $\pi-\pi'$
phase boundaries in fig. 2 (it is not shown for the purposes of clarity).

\section{Slave Boson Mean Field Theory}
We consider an extension of the model of eqn. (\ref{hamil}) by introducing
a flavor -1 $m$ which runs from $1$ to $ {N_ {\rm f}}$.  The Hamiltonian is then
\begin{eqnarray}
{\cal K}&=&\sum_\alpha\sum_{m=1}^ {N_ {\rm f}}\sum_ {\mib q} \left\{
\sum_{\sigma= \uparrow, \downarrow} (\xi ^{\vphantom{\dagger}}_{{\mib q}\alpha}
+ {\mu_{\scriptscriptstyle  {\rm B}}}B \sigma)\,
\psi ^\dagger_{ {\mib q}\alpha m\sigma} \psi ^{\vphantom{\dagger}}_{{\mib q}
\alpha m\sigma}+\right.
\nonumber\\
&&\left.\Delta ^{\vphantom{\dagger}}_\alpha\Big( e^{i\delta_\alpha}\,
\psi ^\dagger_{ {\mib q}\alpha m \uparrow }
\psi ^\dagger_{- {\mib q}\alpha m  \downarrow} + e^{-i\delta_\alpha}\,
\psi ^{\vphantom{\dagger}}_{- {\mib q}\alpha m  \downarrow}
\psi ^{\vphantom{\dagger}}_{ {\mib q}\alpha m  \uparrow }\Big)\right.
\nonumber\\
&&\left.-{1\over \sqrt{ {\cal N}_\alpha}}\sum_{\sigma= \uparrow,
\downarrow}\Big(t ^{\vphantom{\dagger}}_\alpha\,
\psi ^\dagger_{ {\mib q}\alpha m\sigma}c ^{\vphantom{\dagger}}_{m\sigma}
+ t^*_\alpha\, c ^\dagger_{m\sigma}
\psi ^{\vphantom{\dagger}}_{ {\mib q}\alpha m\sigma}\Big) \right\}\nonumber\\
&&+\sum_{m=1}^ {N_ {\rm f}}\sum_{\sigma= \uparrow, \downarrow}
(\varepsilon ^{\vphantom{\dagger}}_0+ {\mu_{\scriptscriptstyle  {\rm B}}}
B\sigma)c ^\dagger_{m\sigma}
c ^{\vphantom{\dagger}}_{m\sigma}\ ,
\end{eqnarray}
where ${\mu_{\scriptscriptstyle{\rm B}}}B$ is the Zeeman energy.  The impurity
level occupancy satisfies the constraint $\sum_{m,\sigma}c ^\dagger_{m\sigma}
c ^{\vphantom{\dagger}}_{m\sigma}\le r$.  We refer to this as model I.
In a slightly different
large-$ {N_ {\rm f}}$ extension (model II), we rescale the hopping amplitudes
$t_\alpha\to t_\alpha/\sqrt{ {N_ {\rm f}}}$ and write the constraint as
$\sum_{m,\sigma}c ^\dagger_{m\sigma}c ^{\vphantom{\dagger}}_{m\sigma} \le r
{N_ {\rm f}}$.  In both cases, $0\le r\le 2$.

Introducing a slave boson $b$ and a Lagrange multiplier $\lambda$ to impose
the constraint, we evaluate the impurity contribution to the free energy at
the mean field level, assuming both $b$ and $\lambda$ to be static.  The
impurity free energy per flavor is then
\begin{eqnarray}
F_{\rm imp}/ {N_ {\rm f}}&=& \varepsilon- {\mu_{\scriptscriptstyle
{\rm B}}}B + p( \varepsilon-  \varepsilon ^{\vphantom{\dagger}}_0)(|b|^2-r)
\nonumber\\
&&\quad+\int\limits_{-\infty}^\infty\!\!\!{d\omega\over\pi}\,
f(\omega+ {\mu_{\scriptscriptstyle  {\rm B}}}B)\,{\rm Im}\,
\ln H(\omega+i 0^+)\nonumber\\
H(\omega)&=&\omega^2- \varepsilon^2-(2 |b|^2
\Gamma ^{\vphantom{\dagger}}_ {\rm a})^2+
{\Big[2|b|^2 \Gamma ^{\vphantom{\dagger}}_ {\rm g}\Delta\sin(\frac{1}{2}\delta)
\Big]^2\over\Delta^2-\omega^2}\nonumber\\
&&\qquad+{4|b|^2  \Gamma ^{\vphantom{\dagger}}_ {\rm a}\,
\omega^2\over\sqrt{\Delta^2-\omega^2}}
\end{eqnarray}
where $ \Gamma ^{\vphantom{\dagger}}_ {\rm a}= \frac{1}{2}$,
$ \Gamma ^{\vphantom{\dagger}}_ {\rm g}=
\sqrt{\Gamma^{\vphantom{\dagger}}_ {\rm L}
\Gamma ^{\vphantom{\dagger}}_ {\rm R}}$ with $\Gamma_\alpha=
\pi\rho_\alpha |t_\alpha|^2$ ($\rho_\alpha$ is the bare density of states per
unit cell in electrode $\alpha$), $\delta=
\delta ^{\vphantom{\dagger}}_ {\rm L}-\delta ^{\vphantom{\dagger}}_ {\rm R}$
is the phase difference between the two superconducting electrodes (we assume
$\Delta_1=\Delta_2\equiv\Delta$), and $f(\omega)=[\exp(\omega/T)+1]^{-1}$ is
the Fermi function.  The Lagrange multiplier $\lambda$ is absorbed into the
renormalized impurity level energy, $ \varepsilon\equiv 
\varepsilon ^{\vphantom{\dagger}}_0+\lambda$.  The factor $p$ in
the second term is $1/ {N_ {\rm f}}$ in model I, while $p=1$ in model II.
Hence it is model II which generates a true large-$ {N_ {\rm f}}$ expansion,u
with all terms in the impurity free energy $F_{\rm imp}$ proportional to
${N_ {\rm f}}$.

From $H(0)<0$, $H(\Delta^-)=+\infty$, and $dH/d\omega>0$, we conclude
that there is a unique solution to the equation $H(\omega)=0$ on the
interval $\omega\in[0,\Delta]$.  Call this root $\Omega$.  We use the
external field $B$ as a Lagrange multiplier so that we may fix the total
value of $S^z$.  Making a Legendre transformation to
$G ^{\vphantom{\dagger}}_{\rm imp}(S^z)=
F ^{\vphantom{\dagger}}_{\rm imp}(B)-2 {\mu_{\scriptscriptstyle  {\rm B}}}
B S^z$, with $S^z=- \frac{1}{2},0,+ \frac{1}{2}$, we set
$ {\partial} F ^{\vphantom{\dagger}}_{\rm imp}/ {\partial}
({\mu_{\scriptscriptstyle  {\rm B}}}B)=0$, and obtain,
at $T=0$ (assuming
$0 <  {\mu_{\scriptscriptstyle  {\rm B}}}B < \Delta$),
\begin{eqnarray}
G^0_{\rm imp}/ {N_ {\rm f}}&=& \varepsilon+p( \varepsilon-
\varepsilon ^{\vphantom{\dagger}}_0)(|b|^2-r) +  {\cal A} \\
{\cal A}&=&{1\over\pi}\int_\Delta^{\omega_\circ}\!\!\!
d\omega\,{\rm Im}\,\ln H(\omega-i0^+)-\Omega\,\delta_{S^z,0}\ ,
\end{eqnarray}
where $\omega_\circ=\sqrt{W^2+\Delta^2}$; $W$ is the half-bandwidth in the
electrodes.  The mean field equations, obtained by setting
$ {\partial}  {G^0_{\rm imp}}/ {\partial} |b|^2=0$ and $ {\partial}
{G^0_{\rm imp}}/ {\partial}  \varepsilon=0$ are
\begin{eqnarray}
1+p(|b|^2-q)+{ {\partial}  {\cal A}\over {\partial}  \varepsilon}&=&0 \\
p( \varepsilon-  \varepsilon ^{\vphantom{\dagger}}_0)+{ {\partial}
{\cal A}\over {\partial} |b|^2}&=&0\ ,
\end{eqnarray}
and the Josephson current is 
\begin{equation}
I={2e\over\hbar}{ {\partial} {\cal A}\over {\partial}\delta}\ .
\end{equation}
Thus, for $0\le {\mu_{\scriptscriptstyle  {\rm B}}}B<\Omega$,
the ground state has $S^z=0$, while for
$\Omega < | {\mu_{\scriptscriptstyle  {\rm B}}}B| < \Delta$, the ground state
has $S^z=- \frac{1}{2}\,{\rm sgn}(B)$.

For $p=r=1$, we can show that the ground state energies satisfy
$\Delta {G^0_{\rm imp}}\equiv {G^0_{\rm imp}}- {G^0_{\rm imp}}>0$,
which means that the mean field slave boson theory cannot describe the $\pi$ or
$\pi'$ phases.  The energy difference $\Delta {G^0_{\rm imp}}$ is minimized at
$\delta=\pi$.  The value of $\Delta{G^0_{\rm imp}}$ is an increasing function
of the bare impurity level energy $  \varepsilon ^{\vphantom{\dagger}}_0$.
However, there are no solutions to the mean field equations for
$ \varepsilon ^{\vphantom{\dagger}}_0< \varepsilon^{\rm min}_0=
-4\pi^{-1} \Gamma ^{\vphantom{\dagger}}_ {\rm a}\cosh^{-1}(\omega_\circ/\Delta)
-2 \Gamma ^{\vphantom{\dagger}}_ {\rm a}\delta_{S^z,0}$.  Rather,
the endpoint solution $ \varepsilon=|b|^2=0$ holds.  Thus, the best we can do is
$\Delta {G^0_{\rm imp}}=0$, but in this case $ {G^0_{\rm imp}}=
\varepsilon ^{\vphantom{\dagger}}_0$, independent of
$\delta$, and there is no Josephson current.

\begin{figure} [!t]
\centering
\leavevmode
\epsfxsize=8cm
\epsfysize=8cm
\epsfbox[18 144 592 718] {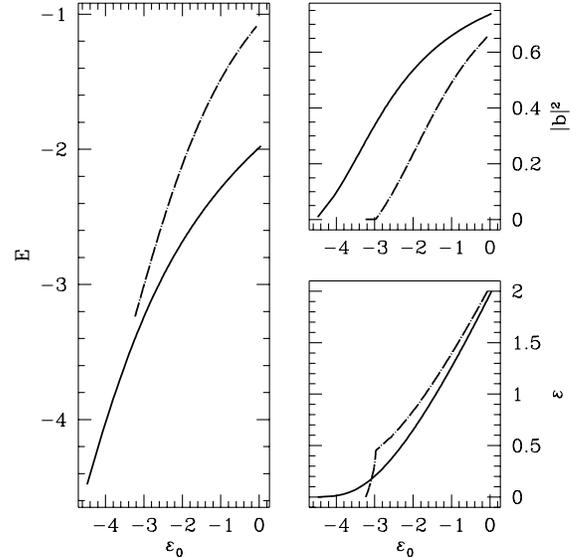}
\caption[]
{\label{cfig} Mean field parameters {\it versus\/}
$\varepsilon ^{\vphantom{\dagger}}_0$. 
$\Delta=1$, $ \Gamma ^{\vphantom{\dagger}}_ {\rm a}=0.650$,
$ \Gamma^{\vphantom{\dagger}}_{\rm g}=0.630$, $B=25$, $\delta=0$, and $p=r=1$.
Solid curve is $S^z=0$, dot-dash is $S^z=\pm \frac{1}{2}$.}
\end{figure}

\begin{figure} [!t]
\centering
\leavevmode
\epsfxsize=8cm
\epsfysize=8cm
\epsfbox[18 144 592 718] {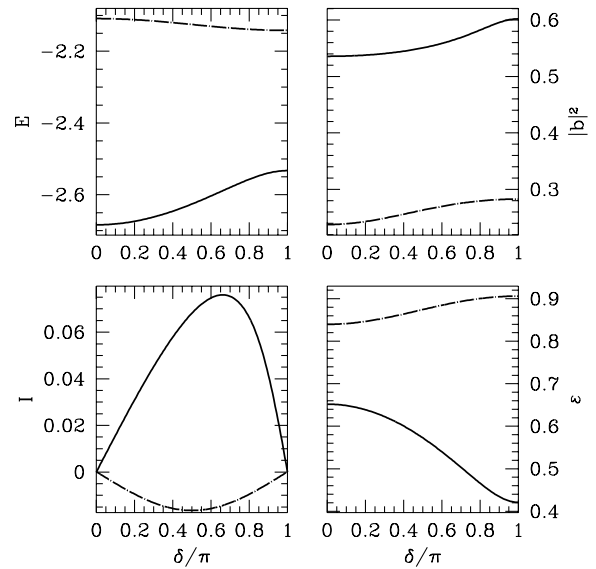}
\caption[]
{\label{dfig} Mean field parameters and Josephson current {\it versus\/}
$\delta=0$ for $  \varepsilon ^{\vphantom{\dagger}}_0=-2$, $\Delta=1$,
$ \Gamma ^{\vphantom{\dagger}}_ {\rm a}=0.650$,
$ \Gamma ^{\vphantom{\dagger}}_ {\rm g}=0.630$, $B=25$, $\delta=0$, $p=r=1$.
Solid curve is $S^z=0$, dot-dash is $S^z=\pm \frac{1}{2}$.}
\end{figure}

\section{Large $\Delta$ Limit}
The case $\Delta\to\infty$ (with $U$ finite) may be solved exactly.
We begin with the Hamiltonian of eqn. (\ref{hamil}), integrating out the
fermion degrees of freedom in the superconductors \cite{RA}.
This generates an induced action,
\begin{eqnarray}
S_{\rm ind}&=&\sum_{\omega_m} {\bar\Psi}_i(\omega_m)\,[\sigma^z\,
{\cal G}(\omega_m)\,\sigma^z]_{ij}\,\Psi_j(\omega_m)\nonumber\\
\Psi(\omega_m)&=&\pmatrix{c ^{\vphantom{\dagger}}_ \uparrow (\omega_m)\cr
{\bar c} ^{\vphantom{\dagger}}_ \downarrow (-\omega_m)\cr}\nonumber\\
\qquad {\cal G} (\omega_m)&=&\sum_\alpha {\Gamma_\alpha\over
\sqrt{\omega_m^2+\Delta_\alpha^2}}\pmatrix{i\omega_m&\Delta
e^{i\delta_\alpha}\cr \Delta e^{-i\delta_\alpha}&i\omega_m\cr}\ .
\end{eqnarray}
If the dynamics all occur on frequency scales $\omega\ll\Delta$, we may
ignore the Matsubara frequencies $\omega_m$ in comparison with $\Delta$.
Adding the induced action to the bare action for the impurity, we find
the resultant action is that for a Hamiltonian
\begin{equation}
{\cal H}_{\rm eff}=  \varepsilon ^{\vphantom{\dagger}}_0
(c ^\dagger_ \uparrow c ^{\vphantom{\dagger}}_ \uparrow +
c ^\dagger_ \downarrow c ^{\vphantom{\dagger}}_ \downarrow )
- {\raise.35ex\hbox{$\chi$}}\,c ^\dagger_ \uparrow c ^\dagger_ \downarrow -
{\raise.35ex\hbox{$\chi$}}^*\,c ^{\vphantom{\dagger}}_ \downarrow
c ^{\vphantom{\dagger}}_ \uparrow + 
U c ^\dagger_ \uparrow c ^\dagger_ \downarrow
c ^{\vphantom{\dagger}}_ \downarrow c ^{\vphantom{\dagger}}_ \uparrow\ ,
\end{equation}
where
\begin{equation}
{\raise.35ex\hbox{$\chi$}}= \Gamma ^{\vphantom{\dagger}}_ {\rm L}
e^{i\delta ^{\vphantom{\dagger}}_ {\rm L}}+
\Gamma^{\vphantom{\dagger}}_{\rm R} e^{i\delta^{\vphantom{\dagger}}_{\rm R}}\ .
\end{equation}
The ground state energy is $E^0_ {\rm D}=  \varepsilon ^{\vphantom{\dagger}}_0$
for the Kramers doublet, and
\begin{equation}
E^0_ {\rm S}=  \varepsilon ^{\vphantom{\dagger}}_0+ \frac{1}{2} U-
\sqrt{(  \varepsilon ^{\vphantom{\dagger}}_0+ \frac{1}{2} U)^2+
| {\raise.35ex\hbox{$\chi$}}|^2}
\end{equation}
for the singlet.  Thus, for $U<U_ {\rm c}$, where
\begin{equation}
U_{\rm c}=4\sqrt{\Gamma_{\rm a}^2-\Gamma_{\rm g}^2\sin^2(\frac{1}{2}\delta)}\ ,
\end{equation}
the Coulomb repulsion is too weak to overcome hybridization effects, and
the ground state is a singlet for all $  \varepsilon ^{\vphantom{\dagger}}_0$.
For $U>U_ {\rm c}$, the ground state will be a doublet provided
\begin{equation}
-U-\sqrt{U^2-U_ {\rm c}^2}<  \varepsilon ^{\vphantom{\dagger}}_0< -U+
\sqrt{U^2-U_ {\rm c}^2}\ .
\end{equation}
This allows the possibility of a level crossing as a function of $\delta$
if $E^0_ {\rm s}<E^0_ {\rm d}<E^0_ {\rm s}$.  However, the doublet
energy is independent of $\delta$ in this model, hence there is no Josephson
coupling in the doublet ground state.

\section{Conclusions}
Using a generalization of the variational wavefunctions applied in the study
of the Kondo effect \cite{VY,GS}, we have demonstrated that the $U=\infty$
interacting impurity Josephson junction exhibits a first order phase transition
for $\Delta\approx T ^{\vphantom{\dagger}}_ {\rm K}$.  When $\Delta
\,{\raise.3ex\hbox{$<$\kern-.75em\lower1ex\hbox{$\sim$}}}\,
T ^{\vphantom{\dagger}}_ {\rm K}$, the ground
state of the system is a singlet, and $E(\delta)$ is minimized at $\delta=0$
and maximized at $\delta=\pi$.  When $\Delta \,{\raise.3ex
\hbox{$>$\kern-.75em\lower1ex\hbox{$\sim$}}}\, T^{\vphantom{\dagger}}_{\rm K}$,
the ground state is a Kramers doublet, the stability of $\delta=0$ and
$\delta=\pi$ is reversed, and the system forms a $\pi$-junction.
For $\Delta\simeq T ^{\vphantom{\dagger}}_ {\rm K}$,
the singlet and doublet energy curves may cross, in which case the ground
state energy has a kink as a function of $\delta$ \cite{RA} and is strongly
nonsinusoidal.

We also solved for the junction's properties within a slave boson
mean field theory.  Unfortunately, this approach is unable to identify
a phase transition, and the singlet state always is lower in energy than
the doublet.  One must go beyond the static mean field approach, as has
recently been accomplished in ref. \cite{CA}, to see the transition.


\section{Bibliography}

\begin{references}

\bibitem{kulik} I. O. Kulik, {\sl Sov. Physics JETP}, {\bf 22}, 841 (1966).

\bibitem{SS} H. Shiba and T. Soda, {\sl Prog. Theor. Phys.} {\bf 41}, 25
(1969).

\bibitem{BKS} L. N. Bulaevskii, V. V. Kuzii, and A. A. Sobyanin,
{\sl JETP Lett.} {\bf 25}, 290 (1977)

\bibitem{GM} L. I. Glazman and K. A. Matveev, {\sl Pis'ma Zh. Eksp. Teor.
Fiz.} {\bf 49}, 570 (1989) [{\sl JETP Lett.} {\bf 49}, 659 (1989)].

\bibitem{SK} B. I. Spivak and S. A. Kivelson, {\sl Phys. Rev. B}
{\bf 43}, 3740 (1991).

\bibitem{RA} A. V. Rozhkov and D. P. Arovas, {\sl Phys. Rev. Lett.}
{\bf 82}, 2788 (1999).

\bibitem{CA}  A. Clerk and V. Ambegaokar, preprint {\tt cond-mat 9910201}.

\bibitem{VY} C. M. Varma and Y. Yafet, {\sl Phys. Rev. B} {\bf 13}, 2950 (1976).

\bibitem{GS} O. Gunnarsson and K. Schonhammer, {\sl Phys. Rev. B} {\bf 28},
4315 (1983).




\end{references}
\end{document}